\documentclass[amsmath,amssymb,prl,aps,superscriptaddress,
floatfix,tightenlines,showpacs,twocolumn
]{revtex4-2}
\usepackage[T1]{fontenc}
\usepackage[latin9]{inputenc}
\usepackage{babel}
\usepackage{calc}
\usepackage{amsmath}
\usepackage{amssymb}
\usepackage{braket}
\usepackage{graphicx}
\usepackage{subfigure}
\usepackage{xcolor}
\usepackage[unicode=true]
 {hyperref}

\def\ket#1{|#1\rangle}

\def\be{\begin{equation}}       \def\ee{\end{equation}}
\def\bea{\begin{eqnarray}}      \def\eea{\end{eqnarray}}
\def\bp{\begin{pmatrix}} \def\ep{\end{pmatrix}}
\def\beaa{\begin{equation}\begin{aligned}}
\def\eeaa{\end{aligned}\end{equation}}

\def\nn{\nonumber}

\makeatletter

\providecommand{\LyX}{\texorpdfstring
  {L\kern-.1667em\lower.25em\hbox{Y}\kern-.125emX\@}
  {LyX}}

\makeatother

\begin{document}
\preprint{APS/123-QED}
\title{Electric Space-time  Translation
and Floquet-Bloch Wavefunction
}

\author{Chenhang Ke}
\affiliation{Department of Physics, Fudan University, Shanghai, 200433, China }
\affiliation{Institute of Natural Sciences, Westlake Institute for Advanced Study, Hangzhou 310024, Zhejiang, China}
\affiliation{Department of Physics, Westlake University, Hangzhou 310024, Zhejiang, China}
\author{Kun Yang}
\affiliation{Department of Physics and the National High Magnetic Field Laboratory, 
Florida State University, Tallahassee, Florida 32306}
\author{Congjun Wu}
\email{wucongjun@westlake.edu.cn}
\affiliation{New Cornerstone Science Laboratory, Department of Physics, School of Science, Westlake University, Hangzhou 310024, Zhejiang, China}
\affiliation{Institute for Theoretical Sciences, Westlake University, Hangzhou 310024, Zhejiang, China}
\affiliation{Key Laboratory for Quantum Materials of Zhejiang Province, School of Science, Westlake University, Hangzhou 310024, Zhejiang, China}
\affiliation{Institute of Natural Sciences, Westlake Institute for Advanced Study, Hangzhou 310024, Zhejiang, China}

\begin{abstract}
As for the study of Landau level wavefunctions for the quantum Hall effect, the magnetic Bloch wavefunctions based on the magnetic translation symmetry have been extensively investigated in the past few decades. 
In this article, the electric Floquet-Bloch wavefunctions based on the electric translation symmetry are
studied as well as the momentum-frequency Brillouin zone, which is applied to the problem of
one dimensional tight-binding model under an external electric field. 
The spectrum of electric Floquet-Bloch states can be generated by the projective representation of electric translation group, and the topological properties of these states are investigated. 
\end{abstract}
                            
\maketitle
The discovery of the quantum Hall effect (QHE) \cite{Klitzing1980} 
has sparked significant interest in the topological properties of electronic wavefunctions \cite{Laughlin1981,TKNN1982}.
In experimental observations, the 2D Hall conductance become quantized into distinct plateaus as varying the applied magnetic field.
The theoretical exploration of the QHE began with the analysis of Landau level
wavefunctions under an external magnetic field, whose spectrum exhibit 
equally spaced flat bands. 
Remarkably, the Hall conductance is quantized in relation to the Chern number \cite{Laughlin1981,TKNN1982,Thouless1983,Avron1985,Niu1985,Bellissard1994,Frohlich1991}.
Since then, extensive efforts have been paid to uncover and classify novel topological phenomenon in condensed matter physics \cite{hasan-kane2010,Qi-Zhang2011}. 
 
Among various theoretical techniques, group-theoretical methods provide profound insights.
The introduction of magnetic translation group \cite{Zak1964} addresses the apparent inconsistency in which the gauge potential appears to break translational symmetry, despite the system being invariant under a physical translation. 
The algebraic structure of the magnetic translations leads to the flatness of Landau levels. 
Moreover, the topological properties are embedded within the representation 
wavefunctions of the magnetic translation group. 
The lowest Landau level wavefunctions, which are variants of the Jacobi 
$\Theta$-function \cite{Haldane2018}, exhibit a single zero in one magnetic unit cell.
As the momentum moves along a non-trivial loop across the magnetic Brillouin zone, the trajectory of the zero also across the magnetic unit cell in real space with the same winding number, indicating a non-trivial Chern number \cite{Hatsugai1993,Kohmoto1989}.

In recent years, non-equilibrium quantum dynamical systems \cite{Oka2019} have attracted a great deal of attentions both in condensed matter \cite{zhu2018,mahmood2016} and cold atom physics \cite{Eckardt2017,Goldman2014}. 
The concept of space-time group has been proposed to describe symmetry properties of a dynamic system \cite{Xu_Wu_2018}. 
The spacial unit cell is extended to 
space-time unit cell, and the Brillouin zone(BZ) is generalized to momentum-frequency BZ.
Such systems exhibit different characteristics in various aspects, including Bloch oscillations \cite{Niu2021}, dynamical topological phenomena \cite{Yang2022,P.Yang2023},
and projective space-time symmetries \cite{Zhang2023} in 1+1 dimensions.

\begin{table*}
\renewcommand\arraystretch{1.5}
\begin{tabular}{|l|l|l|}
\hline & Magnetic translation & Electric translation \\
\hline & $T_{\Delta x}=e^{-\Delta x \partial_x  }$& $T_{\Delta t}=e^{-\Delta t \partial_t  }$\\
Operators & $T_{\Delta y}=e^{- \Delta y \partial_y  } e^{-i e B \Delta y x / \hbar}$& $T_{\Delta x}=e^{- \Delta x \partial_x  } e^{i eE \Delta x t / \hbar}$\\
& {$\left[T_{\Delta x}, H\right]=\left[T_{\Delta y}, H \right]=0$} & {$\left[T_{\Delta x}, i\hbar \partial_t-H\right]=\left[T_{\Delta t}, i \hbar \partial_t-H\right]=0$}\\
\hline Quantization & $e B \Delta x \Delta y=2 \pi \hbar$ & $e E \Delta x \Delta t=2 \pi \hbar$ \\
\hline
Gauge & $\Vec{A} = (-By,0,0),\phi = 0$ &$\Vec{A} = 0,\phi = Ex$ \\
\hline
 \end{tabular}
\caption{This table summarizes the properties of magnetic/electric translation operators. }
\label{Operators}
\end{table*}

Applying an electric field into the spatial lattice introduces electric flux in the space-time domain, similar to the effect of a magnetic field in space. 
The electric field should be described
through a gauge potential, necessitating a specific gauge choice for practical applications.
A conventional choice involves using a time-independent gauge for scalar potential and time-dependent gauge for vector potential. 
In these gauge fixing, the spatial or temporal translation symmetry becomes implicit. 
This leads to the consideration of an electric counterpart to the magnetic translation group, which captures the symmetry of the lattice under an electric field. 
Furthermore, adopting a time-dependent gauge for the electric field can be interpreted as a dynamic system, thereby falling under the classification of space-time groups \cite{Xu_Wu_2018}.
Therefore, the eigen states are dubbed as electric Floquet-Bloch states.

In this article, we investigate the properties of electric Floquet-Bloch states in terms of the electric translation group.
In parallel to the magnetic unit cell and
magnetic BZ, the space-time unit cell and momentum-frequency Brillouin zone
are constructed. 
The 1D tight-binding model in an external electric field is employed as an example to find the exact electric Floquet-Bloch wavefunctions. 
The spectrum of such states in the momentum-frequency Brillouin zone is constructed in terms of the projective representation of the electric translation group.
These wavefunction exhibit a space-time vortex configuration,
which exhibit the Zak phase similar to the Chern insulators. 


We first recall how the magnetic translation is constructed for the quantized motion of a 2D electron in a uniform magnetic field $B$.
The usual translation operator needs to be modified in order to commute with the Hamiltonian $H=(\mathbf{P}-e/c \mathbf{A})^2/2m$. 
Without loss of generality, the Landau gauge is adopted with $A_x=By$ and $A_y=0$.
To render the the translation symmetry explicit,
the translation operators are defined as
$
T_{\Delta x}=e^{- \Delta x \partial_x  },
T_{\Delta y}=e^{- \Delta y \partial_y  } e^{-i e B x \Delta y/ \hbar} ,
$
such that they commute with the Hamiltonian.
Nevertheless, the price to pay is that 
$T_{\Delta x}$ and $T_{\Delta y}$ do not 
commute in general, but satisfies
\beaa
T_{\Delta x} T_{\Delta y}
=e^{i 2\pi \phi/\phi_0 } 
T_{\Delta y} T_{\Delta x},
\eeaa
where $\phi=B\Delta x\Delta y$ is the
flux enclosed by the area spanned by $\Delta x$
and $\Delta y$, and $\phi_0 = hc/e$
is the flux quantum. 
Therefore, $T_{\Delta x}$ and $T_{\Delta y}$
only commute when $\phi=n\phi_0 $
with $n$ an integer. 
To take advantage of the Bloch theorem, 
$\Delta x$ and $\Delta y$ are chosen such that $[T_{\Delta x},T_{\Delta y}]=0$.
Such an area is viewed as the magnetic unit cell for the Landau level problem. 



A similar procedure applies to the system under an electric field to construct the electric translation.
Quantum mechanically, an electric field does not directly enter the Schr\"odinger equation, but via potentials 
$\mathbf{E} =-\nabla \phi -\frac{1}{c} \frac{\partial \mathbf{A}}{\partial t}$.
For simplicity, we can either choose the gauge of using a time-independent scalar potential to generate an electric field, which is dubbed 
{\it ``the static gauge''}, or the one via a time-dependent vector potential, which is dubbed 
``{\it the dynamic gauge}''.
Below we will use the
static gauge, and results of the dynamic gauge is briefly outlined in Supplemental Material (SM) I \cite{supp}.
In such a gauge, the time-dependent Sch\"odinger equation is 
$\left(i\hbar\frac{\partial}{\partial t} - e\phi(x)\right)
\psi =\frac{\mathbf{P}^2}{2m}\psi$,
which is equivalent to a static Hamiltonian as
\bea
H=\frac{\mathbf{P}^2}{2m} + e\phi(x),
\label{eq:static}
\eea
with $\phi(x)=-Ex$.

The electric translation operators are defined as follows
\bea
\label{Electric translation operator}
T_{\Delta t}=e^{- \Delta t \partial_t   }, ~
T_{\Delta x}=e^{- \Delta x \partial_x  } e^{i eE \Delta x t / \hbar}.
\label{eq:elec_tran}
\eea
It is noteworthy that, time translation is considered, emphasis should be placed on the time evolution rather than solely on the Hamiltonian.
The combination $i\hbar \partial_t-H$ is dubbed 
``wave equation operator'', 
and the translation operators defined above commute with the wave equation operator as
\beaa
\left[T_{\Delta x}, i\hbar \partial_t-H\right]=\left[T_{\Delta t}, i\hbar \partial_t-H\right]=0 .
\eeaa

{\it Space-time unit cell and 
momentum-frequency Brillouin zone} ~
$T_{\Delta t}$ and $T_{\Delta x}$ do not 
simply commute but exhibit the algebra relation,
\beaa
T_{\Delta t}T_{\Delta x}=
e^{i2\pi \phi/\phi_0}
T_{\Delta x}T_{\Delta t},
\label{eq:phase}
\eeaa
where the space-time ``electric flux'' is defined as
$\phi=E\Delta x c\Delta t$.
The group generated by 
$T_{\Delta t}$ and $T_{\Delta x}$ 
is dubbed the ``electric translation group''. 
A ``space-time'' unit cell is defined such that the electric flux enclosed by the area spanned by 
$\Delta X$ and $c\Delta t$
is quantized as $\phi=\phi_0$.

For later convenience, we define the space-time unit vectors as
$\mathbf{a_x} = (a_ x, 0), \ \ \
\mathbf{a_t} = (0, 2\pi \hbar/\epsilon(a_ x))
$,
with $\epsilon(a_ x)=eEa_ x$ the potential energy along the field at the distance of $a_ x$.
The space-time electric flux enclosed by the space-time unit cell equals $\phi_0$.
The unit vectors span a 1+1 dimensional (1+1D) discrete space-time lattice, and the space-time translations of Eq. (\ref{eq:elec_tran}) at the lattice vectors form a discrete subgroup in the
space-time domain. 
Correspondingly, the reciprocal lattice vectors in the ``momentum-frequency 
Brillouin zone (MFBZ)'' are represented as 
$\mathbf{K} = (\frac{2\pi}{a_ x}, 0), \mathbf{\Omega} = (0, \frac{\epsilon (a_ x)}{\hbar})$.
The above definitions are summarized in Table (\ref{Operators}).

{\it Electric Floquet-Bloch Wave}
Below  we  discuss a 1D tight-binding model with a uniform electric field $E$ to demonstrate the space-time electric translation with the Hamiltonian defined as
\beaa
H=w \sum_l \left (
c_{l+1}^{\dagger} c_l +h.c.\right )+
\epsilon 
\sum_l l c_l^{\dagger} c_l,
\label{eq:1DHam}
\eeaa
where $\epsilon = eEa$ is the potential drop at one lattice constant; $l$ is the site index; the spin index is omitted for simplicity.
For later convenience, a characteristic frequency is defined as $\hbar \Delta \Omega =\epsilon $. 
The stationary solution with the 
eigenvalue $E_n=n \epsilon$ to the infinitely large system  \cite{Wannier1960,Wilkins1988} is 
\be
\psi_n (l)=J_{n-l}
\left(\frac{2 w}{\epsilon} \right),  
\label{eq:WS_ladder}
\ee
where the energy level index $n$ also denotes the center position of the wavepacket.
The characteristic length associated with this solution is the Bloch oscillation length, defined as  $a_e = \frac{4w}{e E}$, beyond which
$\psi_n(l)$ decays exponentially.
The energy eigenvalues exhibit a tower spectrum as summarized in S. M. II  \cite{supp}. 

The model of Eq. (\ref{eq:1DHam}) serves as a demonstration of a projective representation of the space-time translation group. 
For the lattice Hamiltonian Eq.(\ref{eq:1DHam}), the spatial translation is discrete. 
Now consider the case of total site
number $N$ under the quasi-periodic boundary conditions
$\psi(l,t)=\psi(l+N,t) e^{i N
\epsilon t}$
and $\psi(l,t)=\psi(l,t+T)$ in consistency with Eq.(\ref{eq:elec_tran}).
An integer number of space-time flux quantum are enclosed in the whole system.
The electric translation group $G$ generated by both time  and space translations 
\bea
T_{\Delta t}^p =e^{-p \Delta t \partial_t }, \ \ \,
T_{a}^n=\sum_l 
c_{l+n}^{\dagger} c_l e^{-i n\epsilon t/\hbar  }
\eea
in which $\Delta t = \frac{1}{N} \frac{2\pi}{\Delta  \Omega}$; $n,p$ are integers. 
$\Delta t$ takes a discrete value such that the number of flux quanta enclosed in the space-time area $\Delta t \times Na$ is an integer. 
This representation of the translation group becomes projective due to the additional phase factor given in Eq. (\ref{eq:phase}).

\begin{figure}[h]
\includegraphics[width=0.4\textwidth]{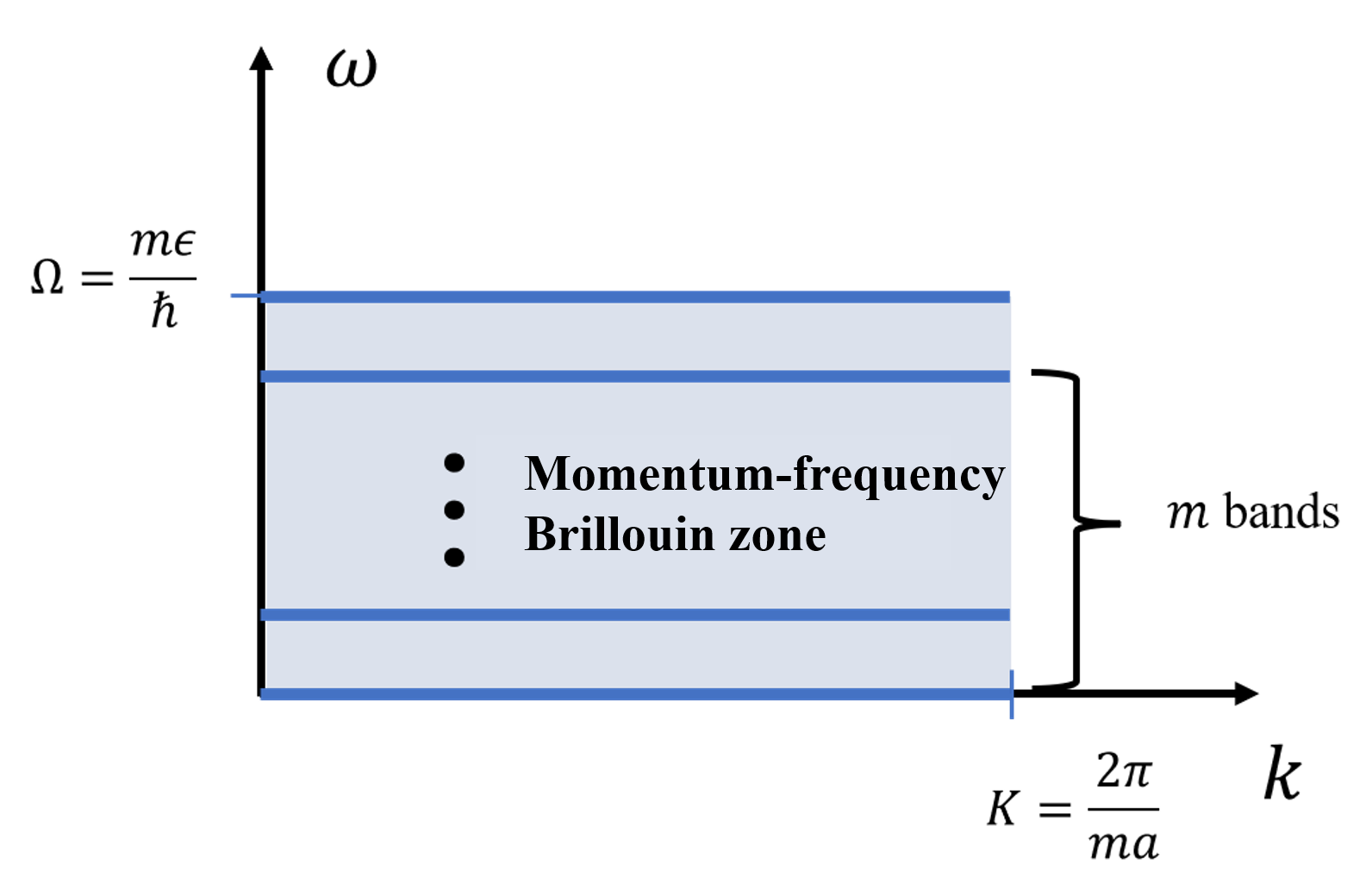}
\caption{This figure presents a schematic depiction of the ``band structure'' for the system. Each deep blue line corresponds to a flat band. These bands are evenly distributed; the ``gap'' between them is   $\epsilon/\hbar$.}
\label{fig:band}
\end{figure}

Given that the electric translation group is non-abelian, we find its Abelian subgroup generated by
two commuting electric translation operators
\bea
T_{a_x}=T_{a}^m, \ \ \, T_{a_t }= T_{\Delta t}^{N/m},
\eea
in which $N/m$ is assumed to be an integer. 
Such an Abelian group defines the space-time unit cell with unit vectors as
$\mathbf{a_x} = (a_ x,0) = (ma,0)$, 
$\mathbf{a_t} = (0, a_ t)$
with $a_ t=N/m \Delta t$.
The momentum-frequency Brillouin zone 
(MFBZ) is defined as depicted in Fig. (\ref{fig:band}) with the reciprocal lattice vectors $\mathbf{K}  = (\frac{2\pi}{ma}, 0)$ and $\mathbf{\Omega} = (0, \frac{m\epsilon}{\hbar})$.  $K = \frac{2\pi}{ma}$ and $\Omega = \frac{m\epsilon}{\hbar}$ are used to denote
the magnitudes of reciprocal lattice vectors.

The common eigenstates of $T_{a_ x}$ and $T_{a_ t}$ are electric Floquet-Bloch wavefunctions defined as
\bea
T_{a_ x} \psi_{k \omega} (l, t) &=&
\psi_{k \omega}(l-m, t) e^{-i eEa_ x\cdot t / \hbar}=
\psi_{k \omega}(l, t) e^{-i k a_ x } , 
\nn \\
T_{a_ t} \psi_{k \omega}(l, t) 
&=&\psi_{k \omega}(l, t-a_ t)
=\psi_{k \omega}(l, t) e^{i \omega a_ t }.
\label{eq:omega}
\eea
Therefore, $\psi_{k,\omega}$'s are characterized by the good quantum numbers of the lattice momentum $k$ and Floquet frequency $\omega$.


For the case of Landau level problem, the magnetic translation symmetry generates the complete flatness of energy spectrum. 
Since $[i\hbar\partial_t - H, G] = 0$, the wavefunction $g\psi_{k,\omega}$, for $g \in G$, also satisfies the Schr\"odinger equation, albeit with altered  momentum and frequency. 
The effect of electric translations to the spectrum
is examined below. 
\bea
T_{a_ t}T^n_a \psi_{k, \omega}& =&e^{-i n \Delta \Omega a_ t  } T_a^n T_{a_ t} \psi_{k,  \omega} 
=e^{i(\omega -n\Delta \Omega) a_ t }
T_a^n \psi_{k, \omega}, \nn \\
T_{a_ x} T_a^n \psi_{k, \omega}
&=&e^{-i k a_ x}
T^n_a \psi_{k, \omega}.
\eea
Hence, the set of good quantum numbers of $T_{a}^n \psi_{k, \omega}$ are
$(k, \omega-n\Delta \Omega)$.
On the other hand, the transformation under a time translation is
\bea
T_{a_ x} T_{\Delta t}^p  \psi_{k, \omega}
& =& e^{i m \Delta K a } T_{\Delta t}^p T_{a_ x} \psi_{k, \omega}
=e^{-i\left(k- \Delta K) a_ x\right.} T_{\Delta t}^p \psi_{k, \omega}, \nn \\
T_{a_ t} T_{\Delta t}^p \psi_{k, \omega}
& =& e^{i p \omega\Delta t } T_{\Delta t}^p 
\psi_{k, \omega},
\eea
where $\Delta K= 2\pi p/N$. 
Hence, $T_{\Delta t} \psi_{k,  \omega}$ is denoted by the set of quantum numbers $(k-\Delta K, \omega)$.

The above results show that if the Schr\"odinger equation has a solution marked by $(k, \omega)$,
then the states of $(k,\omega -\Delta \Omega)$
and $(k-\Delta K, \omega)$ are both solutions. 
This means that the physical states are represented by the points on a discrete grid of $(k,\omega)$ in the MFBZ with the spacings $\Delta K$ and
$\Delta \Omega$ along the momentum and frequency
directions, respectively. 
In other words, $k=t \Delta K$ and $\omega = s \Delta \Omega$ with $1\le t \le N/m$ and 
$1\le s \le m$, respectively. 
Hence, the total number of physical states remain
$N$, i.e., the number of 1D lattice sites.
The number of bands equal $m$, corresponding to the site number in $\mathbf{K}$.
It is also consistent with the dispersion relation illustrated in Fig. \ref{fig:band} as well.
The spectrum flatness and the equal spacing
of frequency $\Delta \Omega$ can also be understood via the freedom of choosing different values of $a_x$
and $a_t$ with a fixed space-time area
as explained in S.M. III \cite{supp}. 

The solutions of the Floquet-Bloch wavefunctions satisfying the boundary conditions of Eq. (\ref{eq:omega})
can be constructed based on Eq. (\ref{eq:WS_ladder}) as
\beaa
\psi_{ k,\omega}(l, t) & =\sum_{q \in \mathcal{Z}} e^{i k\cdot  q a_ x } e^{-i(\omega+q \Omega) t}
J_{l-(\frac{\omega }{\Delta \Omega} +q m)}
\left(\frac{2 w}{\epsilon}\right) \\
 &= \ e^{i k a(l- \frac{\omega}{\Delta \Omega})} e^{-i l \Delta \Omega  t }\\
 &\times \frac{1}{m}\sum_{p=0}^{m-1} 
 e^{-i (l-\frac{\omega}{\Delta \Omega}) \frac{2 \pi p}{m}} e^{i \frac{2 w}{E} \sin \left(\Delta \Omega t  -k a+2\pi \frac{p}{m} \right)},
\label{general_wave_function}
\eeaa
in which 
$J_{l-(\frac{\omega}{\Delta \Omega} +q m)}$ are Bessel functions with its order containing the
lattice site index $l$. 
The second equality in Eq. (\ref{general_wave_function}) arises from the generating function of Bessel functions.
$e^{ix\sin{\theta}} = \sum_{n\in \mathcal{Z}} J_n(x)e^{in\theta}$. The detailed derivation is summarized in the S.M. IV \cite{supp}.

{\it Connection to Bloch oscillations}~
A significant difference between an electron moving in the lattice and the free space is the Bloch oscillation in an electric field.
A wavepacket can be constructed localized both in real and reciprocal spaces for a single band problem, which will propagate and return to its initial configuration exhibiting a periodicity. 
Such a property can be deduced from the tower spectrum described in 
S.M. II \cite{supp}.  
Since the energy eigenvalues are equally spaced with the
interval of $\epsilon$. 
Therefore, the time-evolution of the superpositions of these state exhibit the period of $T_0 = 2\pi/\Delta \Omega$
, which is just the Bloch oscillation perio dicity.  

If the Bloch oscillation wavepackets are constructed by              using wavefunctions of a single band
shown in Fig.\ref{fig:band}.
In this case, only a subset of energy eigenstates in the tower spectrum contributes to the wavepacket, whose
energy eigenvalues form a sub-tower spectrum with 
the energy spacing of $m\epsilon$. 
This indicates that the wave packet that superposed by states in such a band will also demonstrate the Bloch oscillation but with a period $T = T_0/m$.
By taking $m=1$, the ordinary Bloch oscillation is arrived.

{\it Space-time vortex and the Zak phase}~
For a magnetic Bloch wave $\psi_{k_x,k_y}$, it exhibits
a wavefunction vortex around a certain point $(x,y)$ 
in the magnetic unit cell.
Similarly, fixing a point $(x_0,y_0)$ in the magnetic unit cell, $\psi_{k_x,k_y}(x_0,y_0)$ also exhibit a vortex configuration for $(k_x,k_y)$ in the 
magnetic BZ  \cite{Hatsugai1993,Kohmoto1989}. 
As for the lowest Landau level, the magnetic Bloch wavefunctions can be explicitly described by the Jacobi $\theta$-function \cite{Haldane2018}, which exhibit
a single vortex in the magnetic unit cell.
Consequently, the Chern number associated with the lowest Landau level equals one. 

\begin{figure}[h]
\includegraphics[width=\linewidth]{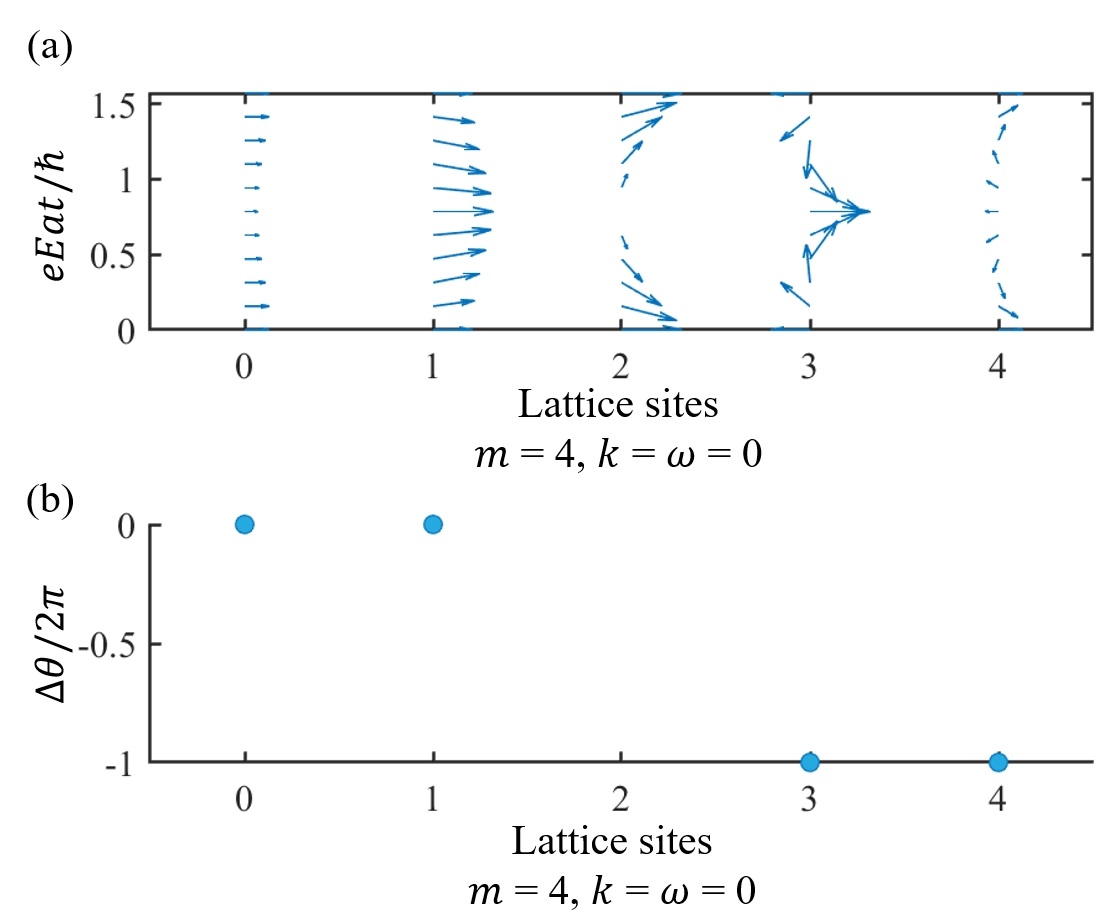}
\caption{($a$) The wavefunction distribution of 
$\psi_{k=0,\omega=0}(l, t)$ is visualized with $m=4$ sites in a space-time unit cell. 
The directions and lengths of arrows represent the phases and magnitudes of the wavefunction.
($b$) For $t\in [0, \frac{2\pi}{m\Delta \Omega})$, the phase winding number
$\Delta \theta/(2\pi)$ along the $t$ direction 
is calculated, which changes from $0$ to $-1$
across the unit cell. 
At $l=2$ where where the wavefunction encounters 
a ``space-time vortex'', the phase winding is not 
well-defined.
}
\label{fig:phasewinding} 
\end{figure}

The wavefunctions  $\psi_{ k,\omega}(l, t)$
of Eq. (\ref{general_wave_function}) obtained from the commuting electric translation operators has a vortex-type structure in the space-time unit cell. 
Based on Eq. \ref{eq:omega}, it can be shown that
the phase winding of $\psi_{k,\omega}$ around 
a space-time unit cell equals $2\pi$, {\it i.e.},
\bea
\int_{t_0}^{t_0+a_t} dt 
\left( \partial_t \theta (l)- 
\partial_t \theta (l+a_l) \right )= 2\pi,
\eea
whose structure is referred as a ``space-time vortex''. 
When $m$ is even, the zeros of the wavefunction are located at $(l,t)$ with 
$l = \frac{\omega}{\Delta \Omega} + \frac{m}{2}$ and $t = \frac{ka}{  \Delta \Omega} + \frac{\pi}{m}
\frac{1}{\Delta \Omega}$.
Consider the case of $\psi_{k,\omega}(l, t)$ when $k = \omega = 0$.
The space-time distribution of the wavefunction
with an even number of sites in the unit cell
is depicted in Fig. \ref{fig:phasewinding}.
As for the case of $m$ taking an odd value, no zero point is guaranteed 
due to the discreteness of lattice site. 
However, the winding number jump can still be found in the wavefunction between 
$l =  \frac{\omega}{\Delta \Omega}+\frac{m+1}{2}$ 
and $l = \frac{\omega}{\Delta \Omega}+\frac{m-1}{2}$. 
The wavefunction zero point can be regarded as hidden between two sites.

The  above ``space-time vortex'' structure results in a non-trivial topological property.
It can be characterized by the Zak phase following
the standard expression,
\beaa
\Theta(\omega=s\Delta \Omega, t)=&\int_0^{\frac{2\pi}{ma}} d k    
\bra{u_{  k,\omega}}  i \partial_k \ket{u_{k,\omega}},
\label{eq:zak}
\eeaa
where $u_{  k,\omega}(l, t)$ is the quasi-periodic kernel defined
in the space-time unit cell via
$ \psi_{  k,\omega}(l, t) = e^{ i ka l} e^{-i  \omega t } u_{k,\omega}(l, t)$
and the Berry connection is defined as
\bea
\bra{u_{  k,\omega}} 
i \partial_k \ket{u_{k,\omega}}= \sum_{l\in\text{unit cell}}  
u_{  k,\omega}^*(l, t) i \partial_k u_{k,\omega}(l, t).
\eea
The Zak phase defined in Eq. (\ref{eq:zak})
is actually time-independent as shown
in S. M. V   \cite{supp}, hence it is denoted as $\Theta(\omega)$, allowing for a simplified representation. 
According to 
$
u_{k,\omega+\Delta \Omega}(l+1, t) = e^{-i k a} u_{k,\omega}(l, t),
$
for every $\omega$, we arrive at
\bea
\Theta((s+1) \Delta \Omega)-
\Theta(s \Delta \Omega)
&=&\int_0^{\Delta \Omega} d k\ \  e^{i k a}  \   i \partial_k  e^{-i k a}  
=\frac{2 \pi}{m}. 
\nn 
\eea
Summing over the discrete values of $\omega$, the 
the increment of $\Theta(\Delta \Omega)$
reaches $2\pi$ when $\omega$ increases by the reciprocal lattice vector $\mathbf{\Omega}$,
\bea
\Theta(\omega +\Omega)-\Theta(\omega)
=2\pi.
\eea

{\it Conclusions}
In summary, the space-time effects of electric translation to the Floquet-Bloch wavefunctions are investigated. 
The momentum-frequency BZ is defined as well as
the space-time unit cell enclosing the quantum
flux of $hc/e$.
Exact solutions to the 1D tight binding model are provided as the quasi-periodic Bloch-Floquet
wavefunctions exhibiting Bloch oscillations.
They possess space-time vortex like structures, and form projective representations of the electric translation group. 
The Zak phase of these states resemble that of quantum anomalous Hall system due to the space-time vortex structure in wavefunctions. 



Note Added:-  
Upon the completion of this manuscript, we became aware of a related work Ref.~\cite{wang2024ssh} 
that studied the same symmetry group in detail and its application in quantum dynamics exhibiting dynamic localization.

\begin{acknowledgments}
We thank J. M. Shi and Z. X. Lin for collaborations on the early stage of this work. 
CW is supported by the National Natural Science Foundation of China under the Grant No. 12234016 and No. 12174317. KY's work is supported by the National Science Foundation Grant DMR-2315954, and performed at the National High Magnetic Field Laboratory which is supported by National Science Foundation Cooperative Agreement No. DMR-2128556, and the State of Florida.
This work has been supported by the New Cornerstone Science Foundation. 
\end{acknowledgments}

\bibliographystyle{prsty}
\bibliography{ref}

\begin{thebibliography}{10}

\bibitem{Klitzing1980}
K.~v. Klitzing, G. Dorda, and M. Pepper, Phys. Rev. Lett. {\bf 45},  494
  (1980).

\bibitem{Laughlin1981}
R.~B. Laughlin, Phys. Rev. B {\bf 23},  5632  (1981).

\bibitem{TKNN1982}
D.~J. Thouless, M. Kohmoto, M.~P. Nightingale, and M. den Nijs, Phys. Rev.
  Lett. {\bf 49},  405  (1982).

\bibitem{Thouless1983}
D.~J. Thouless, Phys. Rev. B {\bf 27},  6083  (1983).

\bibitem{Avron1985}
J.~E. Avron and R. Seiler, Phys. Rev. Lett. {\bf 54},  259  (1985).

\bibitem{Niu1985}
Q. Niu, D.~J. Thouless, and Y.-S. Wu, Phys. Rev. B {\bf 31},  3372  (1985).

\bibitem{Bellissard1994}
J. Bellissard, A. van Elst, and H. Schulz-Baldes, Journal of Mathematical
  Physics {\bf 35},  5373  (1994).

\bibitem{Frohlich1991}
J. Fr\"ohlich and T. Kerler, Nuclear Physics B {\bf 354},  369  (1991).

\bibitem{hasan-kane2010}
M.~Z. Hasan and C.~L. Kane, Rev. Mod. Phys. {\bf 82},  3045  (2010).

\bibitem{Qi-Zhang2011}
X.-L. Qi and S.-C. Zhang, Rev. Mod. Phys. {\bf 83},  1057  (2011).

\bibitem{Zak1964}
J. Zak, Phys. Rev. {\bf 134},  A1602  (1964).

\bibitem{Haldane2018}
F.~D.~M. Haldane, Journal of Mathematical Physics {\bf 59},  081901  (2018).

\bibitem{Hatsugai1993}
Y. Hatsugai, Phys. Rev. Lett. {\bf 71},  3697  (1993).

\bibitem{Kohmoto1989}
M. Kohmoto, Phys. Rev. B {\bf 39},  11943  (1989).

\bibitem{Oka2019}
T. Oka and S. Kitamura, Annual Review of Condensed Matter Physics {\bf 10},
  387  (2019).

\bibitem{zhu2018}
H. Zhu {\it et~al.}, Science {\bf 359},  579  (2018).

\bibitem{mahmood2016}
F. Mahmood {\it et~al.}, Nature Physics {\bf 12},  306  (2016).

\bibitem{Eckardt2017}
A. Eckardt, Rev. Mod. Phys. {\bf 89},  011004  (2017).

\bibitem{Goldman2014}
N. Goldman and J. Dalibard, Phys. Rev. X {\bf 4},  031027  (2014).

\bibitem{Xu_Wu_2018}
S. Xu and C. Wu, Phys. Rev. Lett. {\bf 120},  096401  (2018).

\bibitem{Niu2021}
Q. Gao and Q. Niu, Phys. Rev. Lett. {\bf 127},  036401  (2021).

\bibitem{Yang2022}
Y. Peng, Phys. Rev. Lett. {\bf 128},  186802  (2022).

\bibitem{P.Yang2023}
I. Na {\it et~al.}, Phys. Rev. B {\bf 108},  L180302  (2023).

\bibitem{Zhang2023}
Z. Zhang, Z.~Y. Chen, and Y.~X. Zhao, Communications Physics {\bf 6},  1
  (2023).

\bibitem{supp}
See Supplemental Material at (URL will be inserted by publisher) for additional
  information.

\bibitem{Wannier1960}
G.~H. Wannier, Phys. Rev. {\bf 117},  432  (1960).

\bibitem{Wilkins1988}
J.~H. Davies and J.~W. Wilkins, Phys. Rev. B {\bf 38},  1667  (1988).

\bibitem{wang2024ssh}
J. Wang, J.~J. He, and Q. Niu, Fractional Stark Ladders and Novel Quantum
  Dynamics of Space-time SSH Lattices, 2024.

\end{thebibliography}


\begin{thebibliography}{1}

\bibitem{Yang2019}
S. Girvin and K. Yang, {\em Modern condensed matter physics} (Cambridge
  University Press, ADDRESS, 2019).

\end{thebibliography}

\end{document}


\title{Supplemental Material for ``Space-time Electric Translations"}

\author{Chenhang Ke}
\affiliation{Department of Physics, Fudan University, Shanghai, 200433, China }
\affiliation{Institute of Natural Sciences, Westlake Institute for Advanced Study, Hangzhou 310024, Zhejiang, China}
\affiliation{Department of Physics, Westlake University, Hangzhou 310024, Zhejiang, China}
\author{Kun Yang}
\affiliation{Department of Physics and the National High Magnetic Field Laboratory, 
Florida State University, Tallahassee, Florida 32306}
\author{Congjun Wu}
\email{wucongjun@westlake.edu.cn}
\affiliation{New Cornerstone Science Laboratory, Department of Physics, School of Science, Westlake University, Hangzhou 310024, Zhejiang, China}
\affiliation{Institute for Theoretical Sciences, Westlake University, Hangzhou 310024, Zhejiang, China}
\affiliation{Key Laboratory for Quantum Materials of Zhejiang Province, School of Science, Westlake University, Hangzhou 310024, Zhejiang, China}
\affiliation{Institute of Natural Sciences, Westlake Institute for Advanced Study, Hangzhou 310024, Zhejiang, China}

\pacs{}

\maketitle

\onecolumngrid
\renewcommand{\theequation}{S\arabic{equation}}
\setcounter{equation}{0}
\renewcommand{\thefigure}{S\arabic{figure}}
\setcounter{figure}{0}
\renewcommand{\thetable}{S\arabic{table}}
\setcounter{table}{0}


\tableofcontents

 
\section{I. Electric translation group under the time-dependent gauge }
\label{appnd_A}
In this appendix, we formulate the electric translation group under time-dependent gauge, with the Schr\"odinger equation 

\beaa
    \left[i \hbar \partial_t-\frac{1}{2 m}\left(-i \hbar \partial_x+e E t\right)^2\right] \psi=0.
\eeaa

Now the Hamiltonian is translational invariant in the spacial direction, but not in the temporal direction. 
We should  modify our former definition of electric translation operators in  by

\beaa
T _{\Delta x}=e^{-\Delta x \partial_x}, \quad T _{\Delta t}=e^{-\Delta t \partial t} e^{-i \Delta t e E x/\hbar}.
\eeaa

It can be easily verified that both of them commute with the  `wave equation operator' $i\hbar\partial t-H$ and the exchange of them will cause extra phase

\beaa
T_{\Delta x}T_{\Delta t} =e^{ieE\Delta x \Delta t/\hbar} T_{\Delta x}T_{\Delta t}.
\eeaa

The choice of `space-time unit cell' is identical to that in the main content. We can pick two commuting operators $T_{\delta x}$ and $T_{\delta t}$, with $eE\delta x\delta t=2\pi\hbar$, to generate the abelian space-time translational subgroup.

\section{II. Stationary solution}
\label{Appendix. B}
This section summarizes the stationary solutions of the tight-binding model under the influence of a uniform electric field.
The specific form of the boundary condition is not crucial, as the focus lies on the highly localized nature of these stationary solutions. The Hamiltonians are described as follows:

\be
 H=w \sum_l c_{l+1}^{\dagger} c_l+a E e \sum_l l c_l^{\dagger} c_l+ h.c. 
\ee
Here, the symbol $a$ denotes the lattice constant, $E$ represents the electric field, and $e$ corresponds to the electric charge. 
For ease of expression, we introduce an energy scale, $\epsilon = eEa$, representing the potential energy difference between adjacent lattice sites.

The structure of the energy spectrum can be seen by the commutator of ordinary translation symmetry $\tilde{T}_{ a}=\sum_l c_{l+1}^{\dagger} c_l$ and the Hamiltonian $H$. 
\beaa
[H, \tilde{T}_{a}] =  \epsilon \  \tilde{T}_{ a}   
\eeaa
Suppose we have an energy eigenstate with energy $\lambda$, we can perform the translation operator on it and obtain another eigenstate with energy $\lambda+\epsilon$.
The spectrum should then form an equally spaced tower. 

We can reformulate the Schr\"odinger equation
\be
\psi(l-1)+\psi(l+1)=\frac{\epsilon}{w}\left(\frac{\lambda}{\epsilon}-l\right) \psi(l)
\ee
where $\lambda$ is the energy eigenvalue.
 The Bessel function gives the solution to this equation. 
 Insight into the solution can be derived from the recurrence relation inherent in Bessel functions.
\be
J_{l-1}(x)+J_{l+1}(x)=\frac{2 l}{x} J_l(x)
\ee
    
Upon comparing this equation to the Schr\"odinger equation, we can deduce the form of the stationary solution.

\be
\psi(l)=J_{m-l}\left(2 w/\epsilon\right),  m = \frac{\lambda}{\epsilon} \in \mathcal{Z}
\ee

Only the integer-order Bessel function is selected, as any other choice would lead to an exponential divergence of the wave function.
The characteristic length associated with this solution is the Bloch oscillation length, defined as $L = \frac{4w}{Ee}$.
Beyond this length, the wave function exhibits exponential decay.

\section{III. Magnetic translation group and flatness of Landau level}
\label{Appendix C}

The 2D electron gas under a magnetic field forms Landau level. The flatness of the Landau level can be deduced in a simple and elegant way \cite{Yang2019}. 

As discussed in the main content, the Hamiltonian exhibits magnetic translation symmetry. 
Therefore, to take advantage of the Bloch theorem, the magnetic flux $\phi$ enclosed in the unit cell should be an integer of flux quantum $\phi_0$. 
We assume this integer to be one for simplicity. 
However, the shape of the unit cell is not constrained, given that the magnetic flux is only proportional to the area and the magnetic field. 
In duality, the only constraint on the reciprocal lattice vectors is that the enclosed area should be $(2\pi)^2/\phi$.
Given the energy should be a periodic function of reciprocal lattice, the Landau level must have constant energy all over the Brillouin zone.
 
This argument can be extended to the electric field scenario.  
In the Landau problem, the above argument concludes that the states that differ by any possible reciprocal lattice vector should have the same energy. 
In the electric translation problem, two such states should both be the solution to the schr\"{o}dinger equation. 
A parallel statement can be justified in two limits. Firstly, we pick one lattice constant $a$ in the spatial unit vector ($m=1$) defined in the main content. Then the magnitude of the frequency reciprocal vector is $\Omega = \Delta\Omega$. This indicates a set of equally spaced solutions with interval $\Delta\Omega$ must exist. As well, we can take a very large $m$, then the magnitude of the momentum reciprocal vector should be very small ($K = \frac{2\pi}{ma}$). That means to a solution $\phi_{k,\omega}(l,t)$, $k$ can take a nearly continuous value if the entire system is large enough.

\section{IV. The space-time wave function}

The electric Floquet-Bloch wave functions in the main content are coherently superposed by stationary states, which are Bessel functions.
 \beaa
\psi_{ k,\omega}(l, t) & =\sum_{q \in \mathcal{Z}} e^{i k\cdot  q a_ x } e^{-i(\omega+q \Omega) t}
J_{l-(\frac{\omega }{\Delta \Omega} +q m)}
\left(\frac{2 w}{\epsilon}\right). 
\eeaa
It can be easily verified that these solutions satisfy the Schr\"odinger equation. The infinite summation can be tackled into finite by the generating function of Bessel function.
\beaa
\sum_{n\in\mathcal{Z}} J_{m n+s} e^{i(m n+s) \theta}=\frac{1}{m} \sum_{p=0}^m e^{-i p \cdot s \cdot \frac{2 \pi}{m}} e^{i x \sin \left(\theta+\frac{p}{m} \cdot 2 \pi\right)}.
\eeaa

Therefore, the wavefunction becomes
\beaa
\psi_{ k,\omega}(l, t) 
 = \ e^{i k a(l- \frac{\omega}{\Delta \Omega})} e^{-i l \Delta \Omega  t }
 \times \frac{1}{m}\sum_{p=0}^{m-1} 
 e^{-i (l-\frac{\omega}{\Delta \Omega}) \frac{2 \pi p}{m}} e^{i \frac{2 w}{E} \sin \left(\Delta \Omega t  -k a+2\pi \frac{p}{m} \right)}.
\label{general_wave_function}
\eeaa
Where the frequency and momentum takes  discrete values with spacing $\Delta \Omega$ and $\Delta K$.

\section{V. The proof of the time-independence of the Zak phase} 
In this appendix we give proof of the time-indepence of the Zak phase.
Here we only prove the Zak phase is time-independent for $\omega = 0$.  For generic frequency, the time Independence can be obtained by proof in the main content.

The  quasi-periodical kernel can be subtracted from Floquet-Bloch wavefunctions 
\beaa
 u_{k,0}(l,t) &=e^{-il\Delta\Omega t}  \frac{1}{m}\sum_{p=0}^{m-1} e^{-i l \frac{2 \pi p}{m}} e^{i \frac{2 w}{E} \sin \left(\Delta\Omega t -k a+\frac{2 \pi p}{m}\right)}   \overset{\triangle}{=}f(l,t)g(l,\Delta\Omega t-ka),
\eeaa
where the short handed notation in the second line represents 
\beaa
f(l,t) &= e^{-il\Delta\Omega t} , \ g(l,\Delta\Omega t-ka)&= \frac{1}{m}\sum_{p=0}^{m-1} e^{-i l \frac{2 \pi p}{m}} e^{i \frac{2 w}{E} \sin \left(\Delta\Omega t -k a+\frac{2 \pi p}{m}\right)}.
\eeaa

Firstly, the factor $f(l,t)$ does not contribute to the Berry connection since it cancels with its complex conjugate.
Casting the wave function into Zak phase expression, we arrive
\beaa
\Theta(0,t)&=\int_0^{\frac{2\pi}{ma}} d k \sum_{l\in \text{unit cell}}g^*(l,\Delta\Omega t-ka)\  i\partial k\ g(l,\Delta \Omega t-ka).
\label{eq:zak_w=0}
\eeaa

Because the function $g(l,\Delta\Omega t-ka)$ is a periodic function in $k$, its derivative and its complex conjugation are both periodic function. These three functions have  periodicity $\frac{2\pi}{ma}$ in $k$.  Given that  the integration is over one period, Changing $t$ does not affect the Zak phase. Therefore the Zak phase is time-independent.


\nocite{*}
\bibliographystyle{prsty}
\bibliography{supp.bib}